\documentclass[twocolumn,amsmath,amssymb,floatfix,superscriptaddress,prl,noeprint,aps]{revtex4-2}

\usepackage{bm,graphicx,url,epsf,color}
\usepackage{amssymb}
\usepackage{amsthm}

\usepackage{overpic}
\usepackage{graphicx}
\usepackage{physics}

\usepackage[colorlinks=true,linkcolor=blue,citecolor=blue]{hyperref}

\begin{document}

\preprint{APS/123-QED}

\title{Topological protection of Majorana polaritons in a cavity
}

\author{Zeno Bacciconi}
\affiliation{International School of Advanced Studies (SISSA), via Bonomea 265, 34136 Trieste,
Italy}
\affiliation{The Abdus Salam International Centre for Theoretical Physics (ICTP), Strada Costiera 11, 34151 Trieste, Italy}
\author{Gian Marcello Andolina}%

\affiliation{JEIP, USR 3573 CNRS, Collège de France, PSL Research University, 11 Place Marcelin Berthelot,  F-75321 Paris, France}

\author{Christophe Mora}
\affiliation{Universit\'e Paris Cit\'e, CNRS,  Laboratoire  Mat\'eriaux  et  Ph\'enom\`enes  Quantiques, 75013  Paris,  France}

\date{\today}

\begin{abstract}
Cavity embedding is an emerging paradigm for the control of quantum matter, offering avenues to manipulate electronic states and potentially drive topological phase transitions. In this work, we address the stability of a one-dimensional topological superconducting phase to the vacuum quantum fluctuations brought by a global cavity mode. By employing a quasi-adiabatic analytical approach completed by density matrix renormalization group calculations, we show that the Majorana end modes evolve into composite polaritonic modes while maintaining the topological order intact and robust to disorder. These Majorana polaritons keep their non-abelian exchange properties and protect a twofold exponentially degenerate ground state for an open chain.
\end{abstract}

\maketitle

\textit{Introduction -}  In recent years the possibility of controlling quantum matter by cavity embedding has attracted a lot of attention \cite{cavityqm, BlochReview,Mivehvar2021,VacuumReview}. Strong coupling to cavity vacuum fluctuations has been predicted to affect material properties in many different contexts such as superconductivity \cite{Sentef_sciadv2018,Curtis_prl2019,Schlawin_prl2019}, ferro-electricity \cite{Latini_pnas2021,Ashida_prx2020,Lenk_2022prb} and topology \cite{Ciuti_prb2021,Chiocchetta2021,Dmytruk_commphys2022,Mendez_prr2020,winter_arxiv2023}. It has been shown experimentally that cavity embedding can modify the critical temperature of a charge density wave transition \cite{Fausti_TaS2}, magneto-transport properties \cite{Faist19} and induce the breakdown of topological protection in integer quantum hall transport \cite{Faist22}. In this context, a single-particle electron-photon Chern number was introduced in Ref.~\cite{nguyen_2023electronphoton}.
Addressing topological properties with a global cavity mode is a subtle issue. As a general rule, the robustness of topological properties is ensured by the locality of perturbations. Coupling to a cavity is inherently non-local, and therefore, there is no guarantee that quantum fluctuations preserve topological protection. A contrasting argument in the context of Majorana fermions is that they bear no charge and therefore couple inefficiently to a cavity electric field~\cite{dartiailh2017} (see also Refs.~\cite{trif2012,schmidt2013,dmytruk2015,dmytruk2023} in the context of microwave resonators). Naive expectations relying on the weak effect of vacuum fluctuations of single-mode cavities on extensive quantities \cite{andolina_prb_2019,lenk2022collective,Pilar_quantum2020} should also be taken with care since topological edge states are intrinsically not extensive. 


In this letter, we address this issue by studying a one-dimensional toy model of a topological superconductor~\cite{kitaev2001,chiu2016}, featuring Majorana end states, strongly coupled to a single-mode cavity, and therefore interacting~\cite{hong2013,takahiro2015,fidkowskikitaev2010,fidkowski2011} via long-range forces. We discuss two models for the cavity, either with an electric field~\cite{Faist19,Faist22} or a magnetic field coupling~\cite{ghirri2023ultra,andberger_arxiv2023_chiralcav,tay2023ultrastrong}. Both models respect the fermionic parity $\mathbb{Z}_2$ symmetry of the superconductor~\cite{kitaev2001}. 
Our approach to studying these many-body topological properties is twofold.
We first employ analytical arguments, based on quasi-adiabatic continuation approach~\cite{Alexandradinata_prb2016,Iemini_prl2017}, to establish the resilience of the topological phase to the all-to-all interaction mediated by the cavity mode. 
The edge modes transform into Majorana polaritons~\cite{trif2012} with a light component and are no longer purely fermionic objects. We also perform controlled Density Matrix Renormalization Group (DMRG) numerical simulations~\cite{white_prb_1993,white_prl_1992,schollwock_aop_2011} with a mixed cavity-matter Matrix Product State (MPS) ansatz~\cite{Halati20a,Mendez_prr2020,Passetti_prl2023,bacciconi_arxiv2023} implementing the $\mathbb{Z}_2$ fermionic parity. We identify four markers for topological order~\cite{turner2011,fidkowski2010,Iemini_prl2017}: (i) ground state degeneracy, (ii) entanglement spectrum degeneracy, (iii) non-local edge-edge correlations, and (iv) robustness to local symmetry-preserving perturbations, and demonstrate that they all survive strong cavity quantum fluctuations. We moreover confirm the hybrid nature of the dressed Majorana end operators. Our main finding is that the topological superconducting state is robust to the coupling to the cavity, by adapting its Majorana edge modes, as long as fermionic parity is preserved and no gap closing occurs upon gradually increasing the strength of cavity coupling.

\begin{figure}[t]
\centering
\begin{overpic}[width=0.47\textwidth]{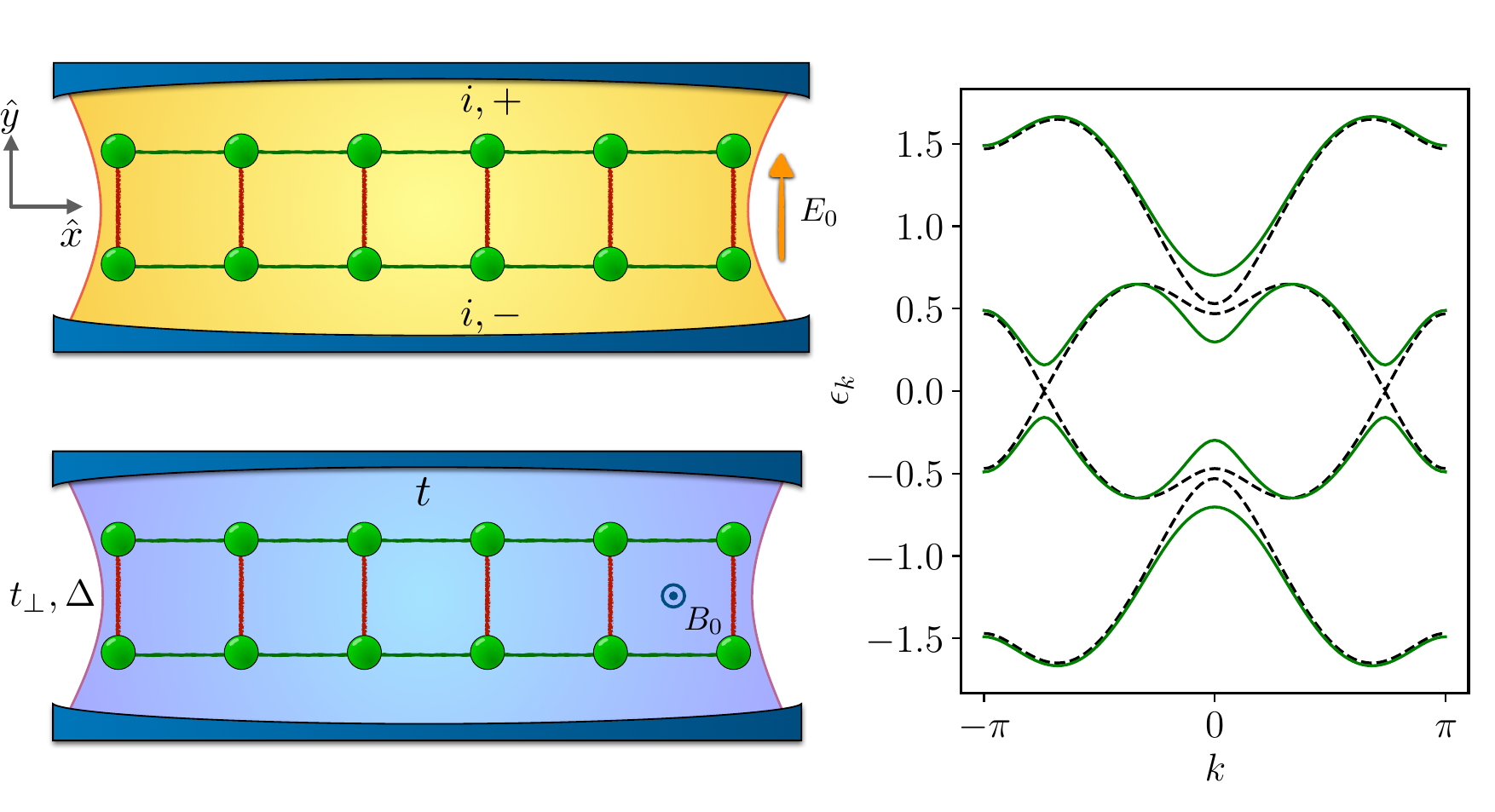}
       \put(-3,52){(a)}
       \put(-3,26){(b)}
       \put(57,52){(c)}
       \end{overpic}
     \caption{Sketch of the two different cavity embeddings. The ladder couples either (a) to a quantized electric field, (b) a quantized magnetic field. (c) Band structure of the cavity-free Hamiltonian $\hat{H}_0$ with (full green line) and without (dashed black line) superconducting pairing.}
         \label{fig:sketch}

\end{figure}

 \textit{The model -}  The starting point of our discussion is a tight-binding model for a one-dimensional topological superconductor. 
 We employ a toy model of spinless electrons hopping on a square ladder geometry \cite{bacciconi_arxiv2023} in the presence of an external magnetic field and a superconducting pairing along the rung of the ladder.
 The Hamiltonian reads:
\begin{align}
   \hat{H}_0 =& -t\sum\limits_{j=1}^{L-1}e^{i\sigma \phi_{ext}/2}   \hat{c}^\dagger_{\sigma,j}\hat{c}_{\sigma,j+1}-t_\perp \sum\limits_{j=1}^L \hat{c}^\dagger_{+,j}\hat{c}_{-,j}\; + \nonumber \\ \label{eq:H0}
   &+\Delta \sum\limits_{j=1}^L \hat{c}^\dagger_{+,j}\hat{c}^\dagger_{-,j}+\mu \sum\limits_{\sigma,j=1}^L \hat c^\dagger_{j,\sigma} \hat c_{j,\sigma}\; +\rm{h.c.}~,
\end{align}
where $\Delta$ is the pairing strength, $\mu$ the chemical potential, $t$ the intraleg hopping, $t_\perp$ the interleg hopping, $\phi_{ext}$ the external magnetic flux per plaquette and $\hat{c}_{j,\sigma}$ annihilates an electron on the leg $\sigma=\pm$ and rung $j=1,..,L$ with $+$ is the top leg and $L$ the number of rungs.
While unconventional, this model can be straightforwardly mapped to the nanowire model~\cite{oreg2010,Lutchyn2010} with strong Rashba spin-orbit coupling and proximity-induced superconductivity, a system that has undergone extensive experimental investigation~\cite{Flensberg2021}. In the ten-fold non-interacting classification~\cite{Ryu2010}, the model Eq.~\eqref{eq:H0} falls into class D, protected only by particle-hole symmetry. It has a $\mathbb{Z}_2$ topological invariant and allows for a topological phase with Majorana end states. However, within a many-body context, the true symmetry protecting the topological phase is fermionic parity.

We now add a single-mode cavity with the bare Hamiltonian $\hat{H}_{c}=\hbar\omega_c \hat{a}^\dagger \hat{a}$. 
In order to draw general conclusions, we examine two distinct physical realizations concerning the vector potential in the cavity: a constant magnetic $(B)$ component along $z$  or a constant electric $(E)$ component along $y$ (Fig \ref{fig:sketch}). The light-matter coupling is achieved through a Peierls substitution~\cite{Luttinger_pr1951,li2020electromagnetic,Schiro20,Cottet_prb2015}, where the hoppings are dressed as~\footnote{See Supplementary Material.}:
\begin{align}
    \mathrm{B} &: \quad \hat{c}^\dagger_{\sigma,j}\hat{c}_{\sigma,j+1} \rightarrow e^{i g_B (\hat{a}+\hat{a}^\dagger)}\hat{c}^\dagger_{\sigma,j}\hat{c}_{\sigma,j+1}\;, \label{dressingB} \\
    \mathrm{E} &: \quad \hat{c}^\dagger_{+,j}\hat{c}_{-,j} \rightarrow e^{i g_E (\hat{a}+\hat{a}^\dagger)}\hat{c}^\dagger_{+,j}\hat{c}_{-,j}\;, \label{dressingE}
\end{align}
depending on the scenario. In the following discussion, when referring to both couplings, we will use $g$ as a combined notation for $g_E$ and $g_B$. The full Hamiltonian is $\hat{H} = \hat{H}_0 + \hat{H}_c$ with either the dressing of Eq.~\eqref{dressingB} or Eq.~\eqref{dressingE}. We are interested in a mesoscopic regime and do not scale $g$ with the system size. Our choice is motivated by the nature of strongly confined cavity modes in nanophotonics, such as split-ring resonators, where there are usually a few, energetically well-separated modes with a significant coupling to the electrons~\cite{Faist22}.

No-go theorems~\cite{andolina_prb_2019,Andolina22} prevent photon condensation, {\it i.e.} a coherent non-zero $\langle \hat{a} \rangle$, for the electric field coupling, whereas $\langle \hat{a} \rangle \ne 0$ can emerge in the magnetic case~\cite{Basko19,guerci_prl_2020,andolina_prb_2020}.
In the latter case, the coherent part of the field simply renormalizes $\phi_{ext}$ and can potentially drive the system out of a topological state or vice-versa. Although very interesting, this effect does not come from quantum fluctuations and can described semiclassically~\cite{bacciconi_arxiv2023}. We henceforth fix $\omega_c=t=t_\perp=-\mu=1$, $\Delta=0.4$ and $\phi_{ext}=0.6889\pi$ such that $\langle \hat{a} \rangle$ remains close to zero and the fermionic chain is in a topological phase.

\begin{figure*}
    \centering
    \begin{overpic}
        [width=\linewidth]{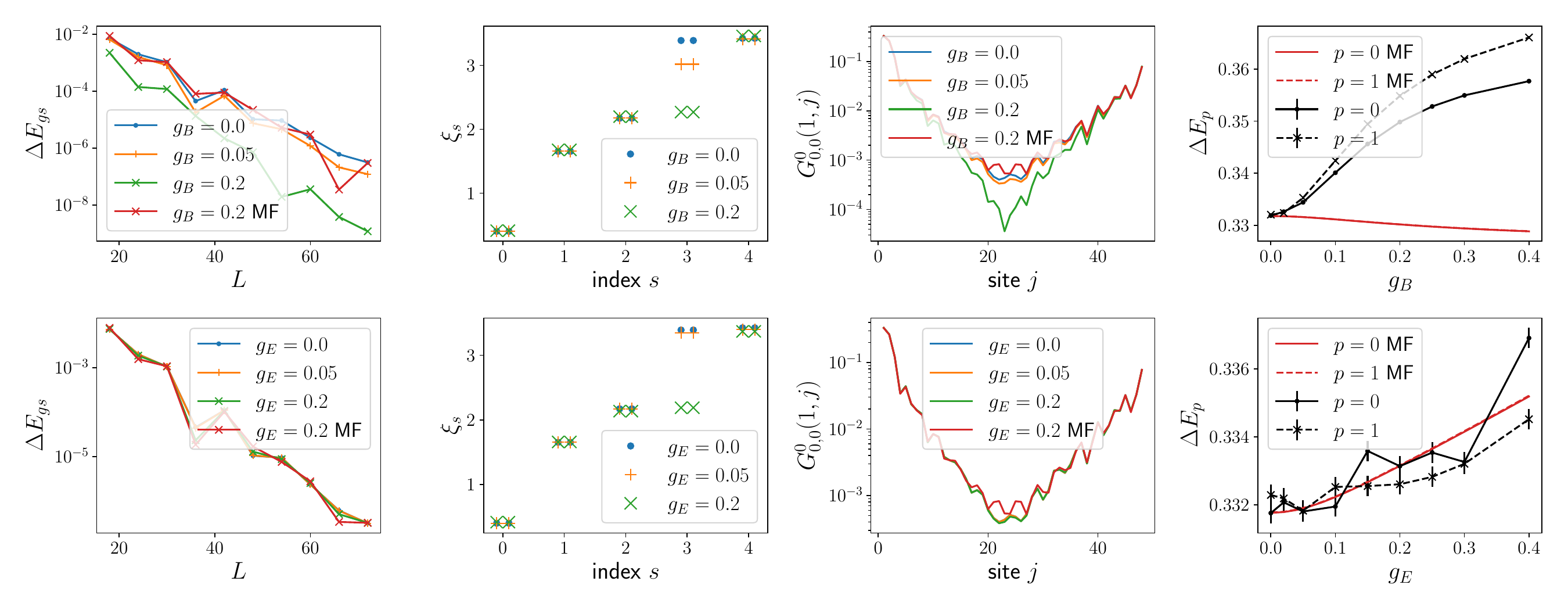}
        \put(21,35.5){(a)}        \put(31,35.5){(b)}
        \put(56,25){(c)}        \put(95,27){(d)}
        \put(6.5,6){(e)}       \put(31,16.5){(f)}
        \put(56,6){(g)}        \put(95,6){(h)}
    \end{overpic}
    \caption{The top (bottom) row shows DMRG results for the magnetic (electric) coupling. (a,e) Ground state energy splitting $\Delta E_{gs}=|E^{gs}_{0}-E^{gs}_1|$     for different coupling strength. The mean-field (MF) result is also shown for comparison. (b,f) Entanglement spectra $\xi_s=-\text{log} \lambda_s$ for an half-chain bipartition.
    The twofold degeneracies come from $(even,even)$ and $(odd,odd)$ parity resolved partitions.
     (c,g) Correlation function for the top leg. 
     (d,h) Energy gap $\Delta E_p$ out of the ground state manifold for both parities, compared with MF (red). The error bars $\sigma_{DMRG}\simeq3\cdot 10^{-4}$ are evaluated from the square root of the discarded weight.      
     $L=48$ except for (a,e)}
    \label{fig:stability}
\end{figure*}
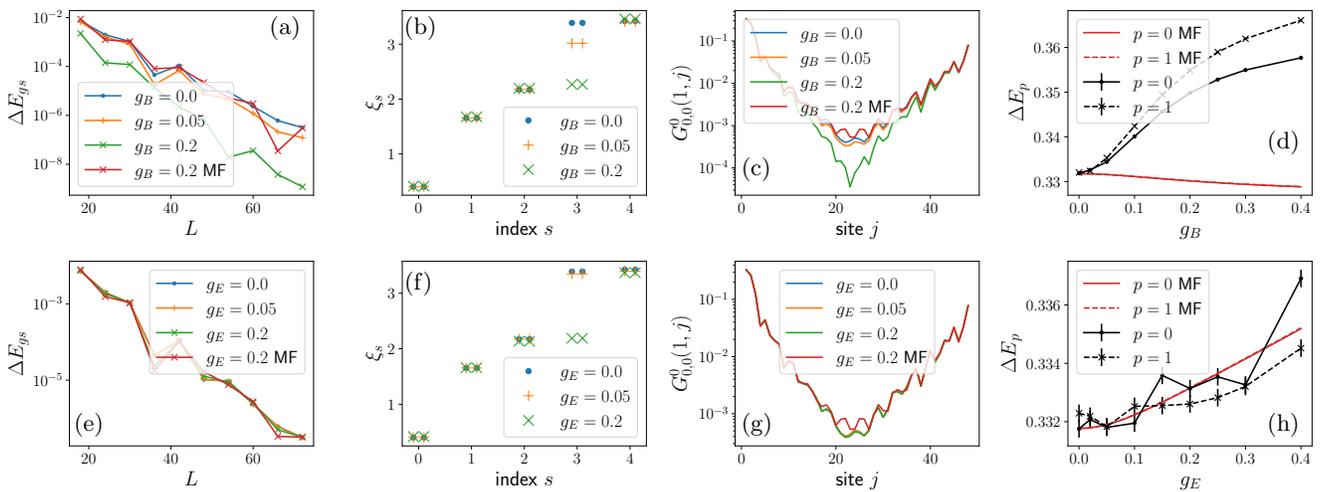

\textit{Majorana dressing -} We first present analytical arguments that elucidate the resilience of topological order to cavity quantum fluctuations.
In the absence of a cavity, the model exhibits Majorana edge modes in its topological phase. The Majorana fermionic operators $\hat{\gamma}_L^0$ and $\hat{\gamma}_R^0$ permute the even and odd parity sectors and protect the ground state (exponential) degeneracy. We employ the theory of quasi-adiabatic continuation~\cite{Alexandradinata_prb2016,Iemini_prl2017} to show that the Majorana operators
\begin{equation}\label{adiabatic}
\hat{\gamma}_L=\mathcal{V}\hat{\gamma}_L^0\mathcal{V}^\dagger\qquad \hat{\gamma}_R=\mathcal{V}\hat{\gamma}_R^0\mathcal{V}^\dagger\:,
\end{equation}
undergo a continuous transformation as the coupling to the cavity is gradually enhanced. The unitary operator $\mathcal{V}$ maps the ground state manifold in the absence of cavity to the one with the cavity.  Importantly, under the assumption that the gradual increase of the cavity coupling maintains both the $\mathbb{Z}_2$ symmetry and a finite spectral gap, it can be shown~\cite{Note1} that $\mathcal{V}$ preserves fermionic locality such that the two deformed Majorana modes remain localized on the two ends of the chain. In addition, $\hat{\gamma}_L$ and $\hat{\gamma}_R$ acquire a finite entanglement with the cavity mode $\hat{a}$ from Eq.~\eqref{adiabatic} and a polaritonic character associated with photonic excitations. They also keep satisfying the Clifford algebra, as $\mathcal{V}_1$ is unitary, and they permute the even- and odd-parity ground states.

Aside from the deformed edge Majorana modes, we also need to prove the persistence of the ground state degeneracy. 
Denoting $P_0$ ($P = \mathcal{V} P_0 \mathcal{V}^\dagger$) the projector onto the ground state manifold without (with) cavity, it is known that~\cite{Alexandradinata_prb2016,kitaev2001}
\begin{equation}\label{locality}
    P_0 \mathcal{O} P_0 = \lambda \, P_0,
\end{equation}
up to exponential corrections with the system size. Here, $\mathcal{O}$ is a local $\mathbb{Z}_2$-symmetric operator and $\lambda$ its eigenvalue in the ground state. Physically, Eq.~\eqref{locality} simply states that a local perturbation cannot distinguish the two ground states because of topological order. Applying Eq.~\eqref{locality} for $\mathcal{O} = \hat{H}_0$ recovers the ground state twofold degeneracy. $\mathcal{V}^\dagger H \mathcal{V}$ is also local for fermions as $\mathcal{V}$ maintains locality. Therefore, we obtain that the even and odd parity states are still exponentially degenerate in the presence of the cavity, or
\begin{equation}
    P \hat{H} P = \mathcal{V} P_0 \mathcal{V}^\dagger \hat{H} \mathcal{V} P_0 \mathcal{V}^\dagger = E_{gs} \mathcal{V} P_0 \mathcal{V}^\dagger = E_{gs} P
\end{equation}
where $E_{gs}$ is the energy of the two ground states. Furthermore, with no cavity, $[\hat{\gamma}^0_\alpha,\hat{H}_0]=0$ (up to exponential corrections), which implies that the twofold degeneracy extends to the whole spectrum and the Majorana operators are called strong edge modes~\cite{Fendley_2012,jermyn2014,Iemini_prl2017}. Such a vanishing commutator is no longer guaranteed in the presence of the cavity. The deformed $\hat{\gamma}_\alpha$ are then weak edge modes as they do not enforce a twofold degeneracy for excited states, only in the ground state manifold.

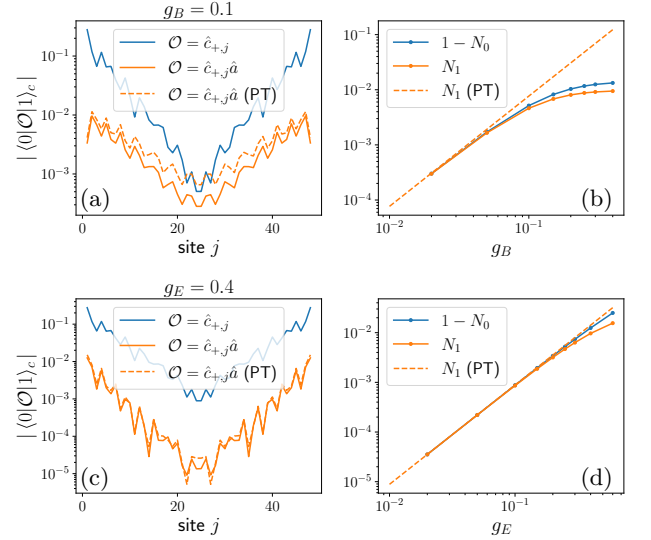
\begin{figure}
    \centering
    \begin{overpic}[width=\linewidth]{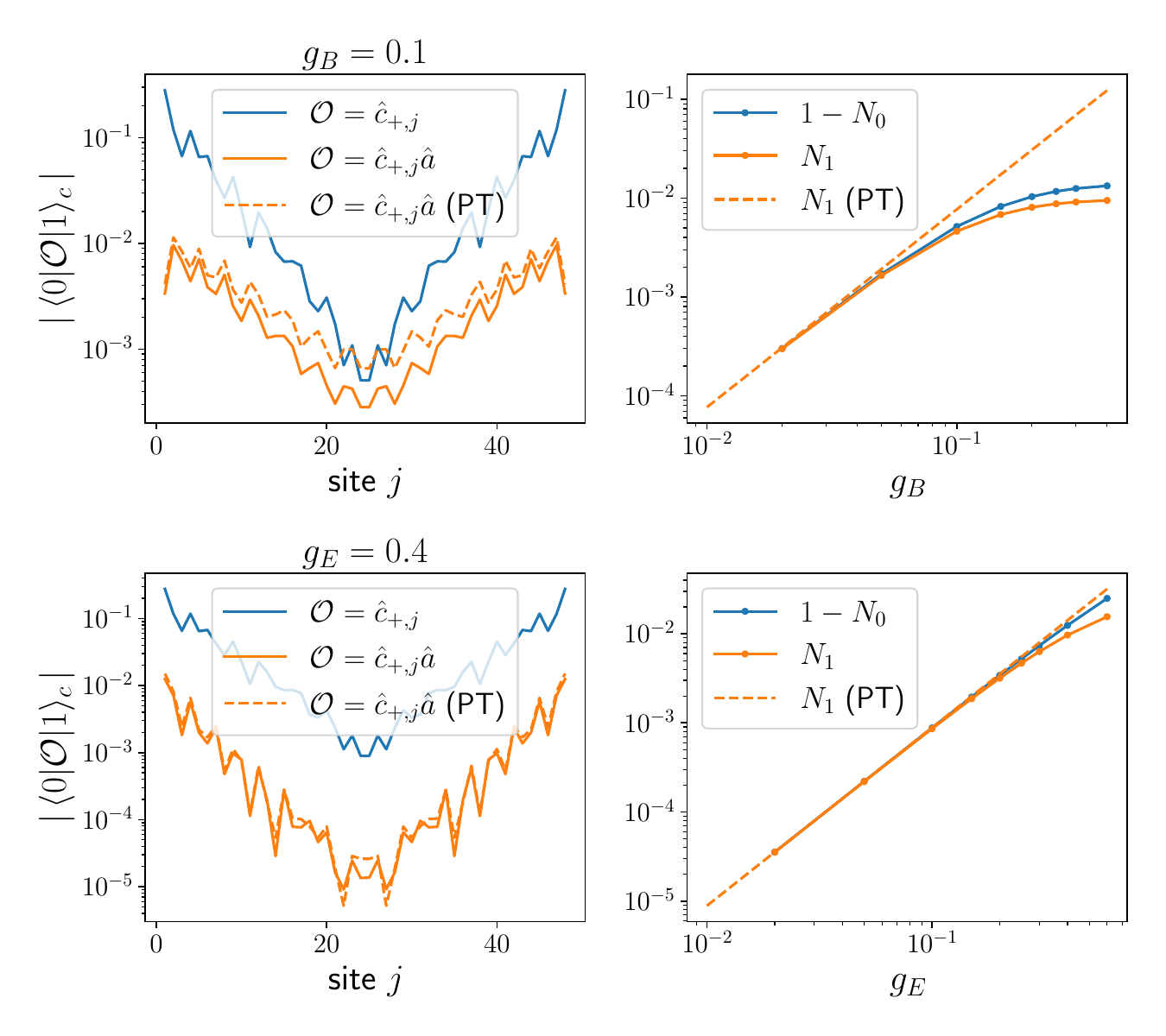}
    \put(13,54){(a)}
    \put(90,54){(b)}
    \put(13,11){(c)}
    \put(90,11){(d)}
    \end{overpic}
    \caption{(a,c) Matrix elements revealing the hybrid nature of Majorana polaritons, with zero ($\mathcal{O}=\hat{c}_{+,j}$) and one photon ($\mathcal{O}=\hat{c}_{+,j} \hat{a}$). (b,d) Evolution of the total zero- and one-photon weights $N_0$ and $N_1$, Eq.~\eqref{eq:weights}, with light-matter coupling for the magnetic (b) and electric (d) cavities. Here $L=48$.}    
    \label{fig:maj_pol}
\end{figure}

Remarkably, the above arguments based on the quasi-adiabatic continuation are very general. They show that any topological superconductor with Majorana end modes is robust to the presence of a cavity, as long the coupling conserves parity ($\mathbb{Z}_2$) and there is an adiabatic path without gap closing to a cavity-free limit. A more explicit polaritonic form can be given to the Majorana fermions from Eq.~\eqref{adiabatic} in perturbation theory
\begin{equation}\label{eq:maj_g}
    \hat{\gamma}_{\alpha} \simeq \hat{\gamma}_{\alpha}^{0} + \sum_{n=(\sigma,j)}\left( \Psi^{1+}_{\alpha}(n) \hat{c}_n^\dagger\hat{a}^\dagger + \Psi^{1-}_{\alpha}(n) \hat{c}_n\hat{a}^\dagger +\rm{h.c.} \right)\;,
\end{equation}
assuming weak coupling to the cavity. The wavefunctions~\cite{Note1} $\Psi^{1+}_{\alpha}(n)$ and $\Psi^{1-}_{\alpha}(n)$ decay exponentially far from the edges as illustrated in Fig.~\ref{fig:maj_pol}(a). 

\textit{Signatures of topology -} In order to test the validity of the discussion above and probe possible non-perturbative effects, we further investigate the Hamiltonian $\hat{H}$ with DMRG calculations. 
We use an hybrid light-matter MPS in which the $\mathbb{Z}_2$ fermionic parity symmetry is implemented~\cite{Note1}, separating the exponentially degenerate even and odd parity sectors. We specifically investigate the four markers, labeled (i)-(iv) in the introduction, as signatures of the topological phase.

The ground state degeneracy $\Delta E_{gs}=|E^{gs}_0 -E^{gs}_1|$ is exponential with the system size up to relatively strong light-matter couplings $g$, confirming point (i). This is reported in Figs. \ref{fig:stability}(a,e) where the ground state energy difference between the two parity sectors is computed. For small $g$, this energy difference can also be evaluated using perturbation theory~\cite{Note1}, confirming its exponential scaling. Interestingly we also report of few ground state parity switches \cite{kitaev2001,majoranawf_Hegde_prb2016} as a function of $g$ (not shown).
The second signature (ii) is the twofold degeneracy in the entanglement spectrum of the the half-chain bipartition. For light-matter systems, one of the two partition has to include the photon. As shown in Fig.~\ref{fig:stability}(b,f) the $g=0$ degeneracy is not broken at finite coupling. The entanglement spectra is nonetheless changing, signalling the presence of finite light-matter entanglement. 

Edge-edge correlations (iii) are shown in Fig.~\ref{fig:stability}(e,g) where the correlator $G_{\sigma,\sigma'}^p(i,j)=\bra{p} \hat{c}^\dagger_{{\sigma},i}\hat{c}_{\sigma',j}\ket{p}$ is calculated on the ground state $\ket{p}$ with parity $p=\{0,1\}$. The revival on the opposite edge reveals the presence of the two end Majorana fermions that both permute the ground states.
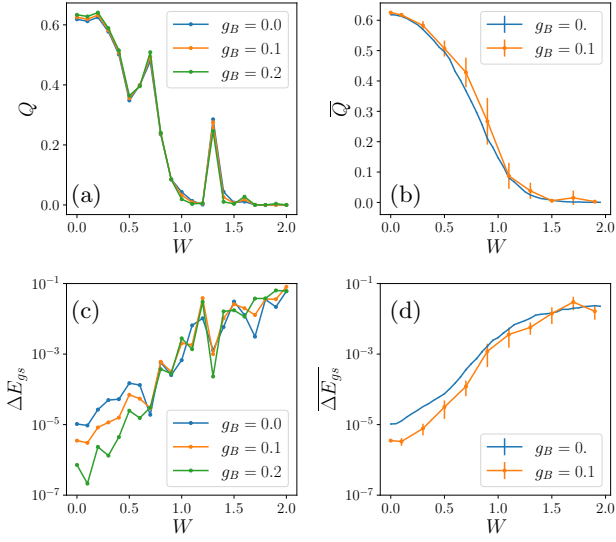
\begin{figure}
    \centering
    \begin{overpic}[width=\linewidth]{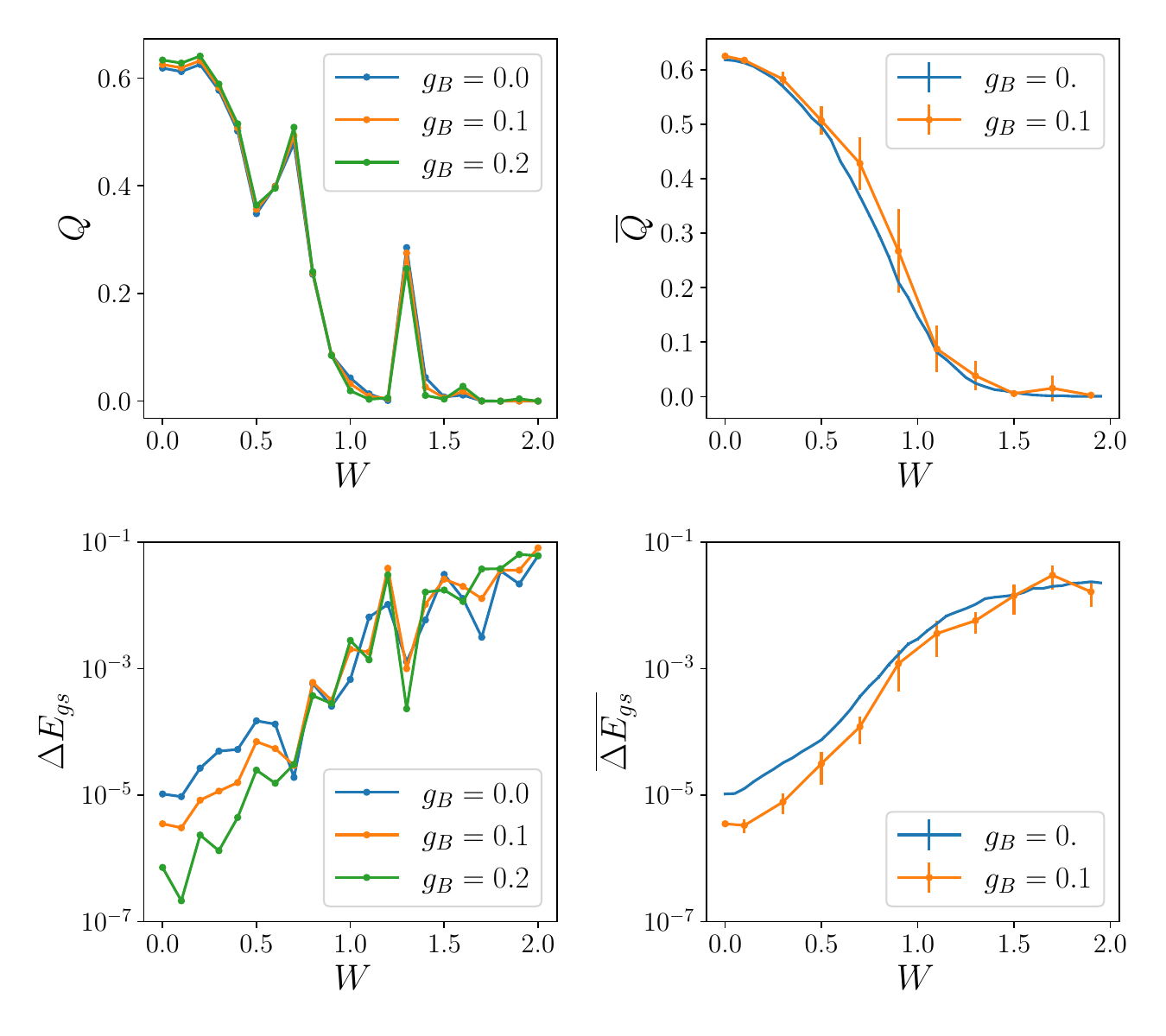}
    \put(13,55){(a)}
    \put(13,37){(c)}
    \put(62,55){(b)}
    \put(62,37){(d)}
    \end{overpic}
    \caption{Magnetic cavity: DMRG results ($g_B>0$) and exact results ($g_B=0$) for topological markers as a function of disorder strength $W$. (a) Edge-edge correlator $Q$ from Eq.~\eqref{defQ} and (c) ground state degeneracy $\Delta E_{gs}$ for a single disorder realization at each strength $W$. The corresponding disorder averaged quantities are shown in (b,d) with $N_\text{dis}=20\,(1000)$ realizations for $g_B=0.1 \,(0.)$. Error bars indicate two standard deviations and $L=48$.}
    \label{fig:disorder}
\end{figure}
We verify (iv) by adding local disorder to the model. Without loss of generality, we consider a gaussian distributed chemical potential, centered around $\mu$, and with the standard deviation $\overline{\delta\mu_{\sigma,i}\delta\mu_{\sigma',j}}=W\delta_{i,j}\delta_{\sigma,\sigma'}$, where $\overline{\mathcal{A}}$ denotes the disorder average of $\mathcal{A}$. As an indicator for edge-edge correlations, we introduce the quantity
\begin{equation}\label{defQ}
    Q= \sum_{\sigma,\sigma'} |G_{\sigma,\sigma'}^0(1,L)-G_{\sigma,\sigma'}^1(1,L)| \;.
\end{equation}
For $g=0$, we have exactly $G_{\sigma,\sigma'}^0(1,L) = - G_{\sigma,\sigma'}^1(1,L)$ as a result of the anticommutation of the two Majorana fermions. In Fig.~\ref{fig:disorder}, we show the indicator $Q$ and the ground state energy splitting $\Delta E_{gs}$ for one disorder realization at each disorder strength (Fig. \ref{fig:disorder}(a,c)) and their disorder average (Fig. \ref{fig:disorder}(b,d) ). The results hardly depend on the strength of cavity coupling, reflecting the robust topological phase. Interestingly, Kohn's theorem yields a different behaviour in the quantum Hall effect, where disorder and cavity collaborate to diminish topological protection~\cite{Ciuti_prb2021,rokaj_arxiv2023,winter_arxiv2023}. Here, disorder plays no role in enhancing the cavity effect on Majorana fermions.

Finally, we test numerically the absence of gap closing as the coupling to the cavity is increased. The gap to the first excited state in each parity sector is shown in Fig.~\ref{fig:stability}(d,h). The splitting of the excited state degeneracy in the magnetic case suggests the evolution from strong to weak edge modes for the Majorana fermions. This could be related to the predicted~\cite{Olesia_prb2023} sensitivity of the cavity damping to the parity of excited electronic states. The comparison with a mean-field approach~\cite{Note1} highlights the need for light-matter entanglement to quantitatively address strong coupling and the many-body nature of the spectrum.

\textit{Majorana polaritons -} As explored through the quasi-adiabatic continuation approach, the Majorana operators $\hat{\gamma}_L$ and $\hat{\gamma}_R$ become increasingly entangled with cavity photons as the light-matter coupling intensifies. This is explicit in Eq.~\eqref{eq:maj_g} in perturbation theory for $g \ll 1$. The polaritonic character of the edge Majorana can be probed with connected matrix elements, such as $\bra{0}\hat{c}_n \hat{a}\ket{1}_c=\bra{0}\hat{c}_n \hat{a}\ket{1}-\bra{0}\hat{a}\ket{0}\bra{0}\hat{c}_n \ket{1}$, which are vanishing at zero coupling $g=0$. To leading order in perturbation theory, we find~\cite{Note1} for instance ($n=(\sigma,j)$):
\begin{align}
    \bra{0}\hat{c}_n\hat{a} \ket{1}_c \simeq \psi^{1+}_L(n)+i\psi^{1+}_R(n)\;,
\end{align}
and other combinations of $\psi^{1\pm}_\alpha (n)$ ($\alpha=L/R$) \footnote{Note that $\psi^{1+}_{\alpha}$ is the part of $\Psi^{1+}_{\alpha}$ with a finite matrix element between the two ground states~\cite{Note1}} are obtained from the connected matrix elements of $\hat{c}_n\hat{a}^\dagger$,$\hat{c}^\dagger_n\hat{a}$ and $\hat{c}_n^\dagger\hat{a}^\dagger$ between $| 0 \rangle$ and $| 1 \rangle$. They are calculated using DMRG, as illustrated in Fig.~\ref{fig:maj_pol}(a,c), and clearly demonstrate localization at the edges as well as photon entanglement. We further quantify the photon mixing by introducing the weights of each component of the Majorana polaritons:
\begin{equation}\label{eq:weights}
    N_0^{\alpha}=\sum_{n}|\psi_{\alpha}^0 |^2 \qquad N_1^{\alpha}= \sum_{n}|\psi_{\alpha}^{1+}|^2 +|\psi_{\alpha}^{1-}|^2 \;,
\end{equation}
where $N_0^R= N_0^L$ and $N_1^R= N_1^L$ by symmetry. $\psi_{\alpha}^0 (n)$ are the purely electronic components of the Majorana operators. The weights calculated from DMRG are shown in Fig.~\ref{fig:maj_pol}(b,d). At small coupling $g$, the missing weight $1-N_0$ from single-fermion contributions is exactly matched by the single-photon polariton sector measured by $N_1$. A deviation between these two lines becomes apparent at strong couplings, indicating additional contributions involving multi-photon and multi-fermion operators. At weak coupling, DMRG aligns well with perturbation theory. 

\textit{Discussion -}  We have shown that a one-dimensional topological superconductor with Majorana end modes is protected against the vacuum quantum fluctuations of an embedding cavity mode, despite its long-range nature. The quasi-adiabatic approach explains this protection and reveals that Majorana evolve into Majorana-polaritons to maintain topological order. We confirm this with DMRG simulations where topological markers are shown to persist through the cavity coupling, and the photonic component of Majorana-polaritons is demonstrated up to strong cavity coupling. The main difference is that the Majorana-polaritons are no longer assured to be strong edge modes. Instead they can transition into weak edge modes where only the ground state is doubly degenerate. 
In the strong cavity coupling regime, mean field techniques prove to be insufficient, making it crucial to account for light-matter entanglement. 
Our argumentation is highly general and is applicable to any 1D phase featuring end Majorana states, regardless of the nature of the cavity coupling, electric or magnetic. Even though not explicitly discussed, the results can be straightforwardly generalized to multi-mode cavities. The crucial prerequisite however is the absence of fermionic parity breaking induced by the cavity coupling. 

Our results suggest that a qubit using Majorana polaritons also requires control over the cavity due to their hybrid nature~\cite{nguyen_2023electronphoton}. We also anticipate that the topological insensitivity to vacuum cavity fluctuations will extend to other topological phases~\cite{shaffer_arxiv2023}, in higher dimensions. For instance, in 2D class $A$ models, such as the quantum Hall effect, the topology is robustly protected by a many-body Chern number~\cite{niu1985}. This does not contradict recent works where finite-size effect and disorder~\cite{Ciuti_prb2021,Arwas_prb2023} or coupling to external degrees of freedom~\cite{rokaj_arxiv2023} have been advocated to predict the loss of conductance quantization observed experimentally~\cite{Faist22}. It would be interesting to build a comprehensive classification of cavity-embedded fermionic models in the spirit of recent classifications of interacting models~\cite{fidkowskikitaev2010,takahiro2015}.

\textit{Acknowledgments- } We acknowledge fruitful discussions with M. Dalmonte, C. Ciuti, T. Chanda, G. Chiriacò, O. Dmytruk and M. Schir{\`o}. The DMRG numerical implementation is done via the ITensor library \cite{itensor}. G.M.A. acknwoledges funding from the European Research Council (ERC) under the European Union's Horizon 2020 research and innovation
programme (Grant agreement No. 101002955 -- CONQUER).

\bibliography{main.bbl}

\clearpage 
\setcounter{section}{0}
\setcounter{equation}{0}%
\setcounter{figure}{0}%
\setcounter{table}{0}%

\setcounter{page}{1}

\renewcommand{\thetable}{S\arabic{table}}
\renewcommand{\theequation}{S\arabic{equation}}
\renewcommand{\thefigure}{S\arabic{figure}}

\onecolumngrid

\begin{center}
\textbf{\Large Supplemental Material for:\\ ``Topological protection of Majorana fermions in a cavity''}
\bigskip

Zeno Bacciconi, $^{1,\,2}$
Gian Marcello Andolina, $^3$ and
Christophe Mora, $^3$
\bigskip
$^1$\!{\it International School of Advanced Studies (SISSA), via Bonomea 265, 34136 Trieste,
Italy}

$^2$\!{\it The Abdus Salam International Centre for Theoretical Physics (ICTP), Strada Costiera 11, 34151 Trieste, Italy}

$^3$\!{\it JEIP, USR 3573 CNRS, Collège de France, PSL Research University, 11 Place Marcelin Berthelot,  F-75321 Paris, France}

$^4$\!{\it Universit\'e Paris Cit\'e, CNRS,  Laboratoire  Mat\'eriaux  et  Ph\'enom\`enes  Quantiques, 75013  Paris,  France}

$^5$\!{\it NEST, Istituto Nanoscienze-CNR and Scuola Normale Superiore, I-56127 Pisa,~Italy}

\bigskip

\end{center}


%
\onecolumngrid
\section{Details on the light-matter coupling}\label{sec:app_lightcoupling}
In the following we detail the Peierls substitution for the tight-binding model under consideration. For general hopping terms the Peierls substitution reads:
\begin{equation}
    \label{Eq:Peierls}
    \hat{c}^\dagger_{i,\sigma} \hat{c}_{j,\sigma^\prime}\rightarrow \exp\Big[{i \int_{R_{i,\sigma}}^{R_{j,\sigma^\prime}}d\bold{r}\cdot \hat{\bold{A}}(\bold{r})} \Big] \hat{c}^\dagger_{i,\sigma} \hat{c}_{j,\sigma^\prime}~,
\end{equation}
with $e=c=\hbar=1$. It is straightforward to show that the electric coupling $\bm{A}_E\propto \hat{y}$ only dresses the vertical hopping $t_\perp$ while the magnetic coupling $\bm{A}_B\propto y \hat{x}$ dresses only the horizontal hopping $t_0$. The superconducting coupling instead is more subtle. Regarding the electric coupling, it has been shown that \cite{Cottet_prb2015}, in order to preserve gauge invariance,  one needs to dress the superconducting pairing terms as 
\begin{equation}    \label{Eq:Peierls_superc}
    \hat{c}_{i,\sigma} \hat{c}_{j,\sigma^\prime}\rightarrow \exp\Big[{i \int_{\bm{R}_0}^{\bm{R}_{i,\sigma}}d\bold{r}\cdot \hat{\bold{A}}(\bold{r})} +{i \int_{\bm{R}_0}^{\bm{R}_{j,\sigma^\prime}}d\bold{r}\cdot \hat{\bold{A}}(\bold{r})} \Big] \hat{c}_{i,\sigma} \hat{c}_{j,\sigma^\prime}~,
\end{equation}
where $\bm{R}_0$ is an arbitrary starting point and $\bm{R}_{j,\sigma}$ the position of the site in rung $j$ and leg $\sigma$. It is easy to check that, with the choice of $\bm{R}_0$ in the center of the ladder, the dressing of the superconducting pairing in equation Eq.~\eqref{eq:H0} in the main text vanishes for $\bm{A}_E\propto \hat{y}$. The orientation choice as well as the choice of the superconducting pairing is made to specifically avoid dressing the superconducting pairing. Equivalently for the magnetic field case with $\bm{A}\propto -y\hat{x}$ we also get a vanishing dressing on the chosen superconducting term.

We now comment on a problem that arises when coupling an electric field along $x$, which would be the natural choice on a Kitaev chain and that we did not include in this work. The dressing would be:
\begin{equation}
   \bm{A}_x\propto\hat{x} \;: \qquad \hat{c}^\dagger_{j,+}\hat{c}^\dagger_{j,-}\rightarrow \hat{c}^\dagger_{j,+}\hat{c}^\dagger_{j,-} e^{2ig_x r_j(\hat{a}+\hat{a}^\dagger)}\;,
\end{equation}
where we introduce a light-matter coupling constant $g_x$ and the rung position $r_j$. This light-matter coupling give rise to some severe numerical issues regarding the truncation of photon Hilbert space as representing a displacement operator $\hat{D}(\alpha)=e^{\alpha \hat{a}-\alpha^*\hat {a}^\dagger}$ requires $N_{ph}\gg \alpha^2$, in the present case would be $N_{ph}\gg r_j^2 \propto L^2$. The ladder geometry reveals here to be crucial, along $y$ we still have freedom to couple non-trivially a cavity mode and we don't get the thermodynamic limit problem of $r_j\propto L$ which instead we get along $x$. \\

\section{Topology in absence of cavity}\label{sec:app_topology}
In the following we analyze the non-interacting Hamiltonian $\hat{H}_0$ in equation \eqref{eq:H0} of the main text and show the presence of a topological phase. In order to do so we introduce momentum space fermions $\hat{c}_{\sigma,k}=\frac{1}{\sqrt{L}}\sum_{j} e^{-ikj}\hat{c}_{\sigma,j}$ and the nambu operator $\hat{\psi}_k =(\hat{c}_{+,k} ,\hat{c}_{+,k},\hat{c}^\dagger_{-,-k},-\hat{c}^\dagger_{+,-k} )$. The hamiltonian can then be written as:
\begin{equation}
    \hat{H}_0 =\frac{1}{2}\sum_k \hat{\psi}^\dagger_k \mathcal{H}_k \hat{\psi}_k
\end{equation}
with $\mathcal{H}_k$ a $4\times4$ matrix:
\begin{align}
    &\mathcal{H}_k= h^A_k \tau_z +h^B_k \sigma_z \tau_z -t_\perp\sigma_x +\Delta \tau_x\;,\\[3mm]
    &h^A_k=-2t\cos(k)\cos(\phi_{ext}/2) \qquad h^B_k = 2t \sin(k) \sin(\phi_{ext}/2)
\end{align}
and $\sigma_{\alpha}$ ($\tau_\alpha$) acting on the leg degree of freedom (Nambu degree of freedom). The model has an intrinsic particle-hole symmetry $P= i\sigma_y\tau_y K$ with $K$ complex conjugation. There is an emergent time-reversal symmetry  $T'=i\sigma_y K$ at $t_\perp=0$ which is broken at finite $t_\perp$. As stated in main text, the topological class is $\mathcal{D}$. The model is characterized by the $\mathbb{Z}_2$ topological index
\begin{equation}\label{topoindex}
    (-1)^\nu = {\rm sgn} \left(t_\perp - \sqrt{\Delta^2 + (h^A_0)^2} \right) {\rm sgn} \left (t_\perp - \sqrt{\Delta^2 + (h^A_\pi)^2} \right)
\end{equation}
given from the Pfaffian~\cite{kitaev2001} at momenta $k=0,\pi$. In the non-trivial phase, $\nu=1$, the finite open ladder exhibits Majorana zero-modes at its boundaries.

More precisely, the model with open boundaries is diagonalized using the Bogoliubov transformation: 
\begin{equation}
    \hat{d}^\dagger_\mu=\sum_{n=(j,\sigma)} u_{n,\mu}\hat{c}^\dagger_n + v_{n,\mu} \hat{c}_n
\qquad \hat{c}^\dagger_{n}= \sum_\mu u_{n,\mu}^*\hat{d}^\dagger_{\mu} +v_{n,\mu} \hat{d}_{\mu}
\end{equation}
where $\sum_n |u_{n,\mu}|^2 + |v_{n,\mu}|^2=1$.
This brings the Bogoliubov-de Genne onto:
\begin{equation}
    H_0= \sum_{\mu=1}^L \epsilon_\mu \hat{d}^\dagger_\mu \hat{d}_\mu 
\end{equation}
The Majorana zero-modes $\gamma_L^0$ and $\gamma_R^0$ will then form the complex fermion $d_1$ with $\epsilon_1\simeq 0$:
\begin{align}    
        &\ket{\overline{1}}=\hat{d}^\dagger_1 \ket{\emptyset} \quad d^\dagger_1= \frac{1}{2}(\gamma^0_L + i \gamma^0_R )= \sum_{n=(\sigma,j)} u_{n,1} \hat{c}^\dagger_n + v_{n,1} \hat{c}_n\\
            &\hat{\gamma}_L^0= \sum_{n} \psi_L^0(n) \hat{c}^\dagger_n + (\psi^0_L(n))^*  \hat{c}_n \qquad \hat{\gamma}_R^0= \sum_{n} \psi_R^0(n) \hat{c}^\dagger_n + (\psi^0_R(n))^*  \hat{c}_n \\
    &\psi^0_L(n) = u_{n,1} + v^*_{n,1} \qquad \psi^0_R(n) = -i(u_{n,1} - v^*_{n,1})\\
    &\sum_n |\psi^0_L(n)|^2 = \sum_n|\psi^0_R(n)|^2  =1
\end{align}
where $\ket{\emptyset}$ is the vacuum 
of quasi-particles and $\ket{\overline{1}}$ the other opposite parity ground state. Note that the fermion number parity of the vacuum $\ket{\emptyset}$ is not always the same and can be shown to be linked to the Pfaffian of the Bogoliubov transformation that diagonalize the single-particle hamiltonian \cite{kitaev2001,majoranawf_Hegde_prb2016}. 

\section{Theory of quasi-adiabatic continuation }\label{sec:app_dressing}

\subsection{General formalism}
We give a more detailed discussion of the quasi-adiabatic dressing procedure. 
The concept here is to introduce an operator that continuously deforms the ground state manifold by progressively increasing the magnitude of cavity fluctuations. In a broader context, it can be applied to implement any perturbation, such as disorder for instance, in a continuous manner. This continuity is achieved by substituting the light-matter coupling constant $g$ with $s g$, where $s$ ranges from $0$ to $1$. This establishes a continuous interpolation denoted as $H_s$ between the model with no cavity fluctuations at $s=0$ and the model with complete fluctuations at $s=1$. The unitary operator $\mathcal{V}_s$ 
\begin{equation}\label{deformation}
\mathcal{V}_s = T e^{i \int_0^s {\mathcal D}_{s'} d s'}   \qquad \qquad {\mathcal D}_{s} = - i \int_{-\infty}^{+\infty} d t \, F(\Delta_s t) \, e^{i H_s t} \partial_s H_s e^{-i H_s t}, 
\end{equation}
maps the ground state manifold at finite $s$ to the one in the absence of cavity quantum fluctuations at $s=0$. $\Delta_s$ is the spectral gap above the ground state subspace. The function $F(t)$ suppresses the long-time evolution to avoid a spatial spreading of operators spoiling locality as discussed below. The Fourier transform of $F(t)$ must satisfy 
\begin{equation}\label{FTF}
    \tilde F(\omega) = \int_{-\infty}^{+\infty} d t e^{i \omega t} F(t) = - \frac{1}{\omega}
\end{equation}
for $|\omega|\ge 1$. The choice of $\mathcal{V}_s$ is not unique (as the mapping of ground states does not fully determine $\mathcal{V}_s$) and, consequently, the form of $\tilde F(\omega)$ is not fixed for $|\omega| <1$. It must however be such that $F(t)$ decays exponentially at long times to preserve locality. We now prove the need for Eq.~\eqref{FTF} following Ref.~\cite{Alexandradinata_prb2016}. We define the eigenstates of $H_s$ as $|j,s\rangle$ with $H_s |j,s\rangle = E_j^s |j,s\rangle$, and the spectrum $E_j^s$ labeled by $s$. Using Eq.~\eqref{FTF}, the matrix elements of ${\mathcal D}_{s}$ are expressed as
\begin{equation}\label{matrix-elements}
    i \langle j,s | {\mathcal D}_{s} |j',s\rangle = - \frac{\langle j,s | \partial_s H_s |j',s\rangle}{E_{j'}^s-E_j^s} = \langle j,s | \partial_s |j',s\rangle,
\end{equation}
where either $j=0,1$ (of the ground states) or $j'=0$ so that $|E_{j'}^s-E_j^s|\ge \Delta_s$. We introduce the projector onto the ground state at finite $s$, $P_s = \sum_{j=0,1} |j,s\rangle \langle j,s |$. Then, we have
\begin{equation}
  \langle j_1,s |  \partial_s P_s |j_2,s\rangle  = \sum_{j=0,1} \Big( \langle j_1,s | \partial_s |j,s\rangle \langle j,s | j_2,s\rangle - \langle j_1,s | j,s\rangle \langle j,s | \partial_s |j_2,s\rangle \Big) = i \langle j_1,s | [ {\mathcal D}_{s} , P_s ] | j_2,s\rangle,
\end{equation}
regardless of $j_1,j_2$ and where we used Eq.~\eqref{matrix-elements}. This proves that $\partial_s P_s = [ {\mathcal D}_{s} , P_s ]$ so that 
\begin{equation}\label{rotated-projector}
    P_s = \mathcal{V}_s P_0 \mathcal{V}_s^\dagger 
\end{equation}
where the solution for $\mathcal{V}_s$ is given in Eq.~\eqref{deformation}. Eq.~\eqref{rotated-projector} clearly identifies $\mathcal{V}_s$ as the unitary operator rotating the ground state manifold, and we have proven here that it requires the Fourier transform given in Eq.~\eqref{FTF} (for $|\omega|\ge 1$).

One crucial point in the theory is the preservation of locality. This can be understood by examining the expression of ${\mathcal D}_{s}$ in Eq.~\eqref{rotated-projector}. One first considers $\partial_s H_s$ which is local (in fact a sum of local terms) for fermionic variables since $H_s$ is. Then $\partial_s H_s (t) = e^{i H_s t} \partial_s H_s e^{-i H_s t}$ evolves this operator in time and the exponential suppression of $F(t)$ ensures that the final result for ${\mathcal D}_{s}$ is local.

\subsection{Perturbation theory}

The results of adiabatic continuity are very powerful and general and they only require the perturbation to maintain the parity symmetry and the gap to remain finite. However, they only give a closed form for the rotated Majorana and an analytical solution is possible only in perturbation theory in $g$. Performing such a perturbation from Eq.~\eqref{rotated-projector} for the magnetic cavity, we find:
\begin{equation}\label{pertur-mf}
    \hat{\gamma}_L\simeq \hat{\gamma}_L^0 + ig_B \left[\mathcal{D}_0,\hat{\gamma}_L^0\right] +\mathcal{O}(g_B^2)
\end{equation}
The Hamiltonian expands as $\hat{H} = \hat{H}_0+V_1 \, (\hat{a}+\hat{a}^\dagger)$ with
\begin{equation}\label{eq:V1_expansion}
    \hat{V}_1= \sum_{\alpha=L,R} C_\alpha \hat{\gamma}^0_\alpha + \dots  \qquad C_\alpha = -\sum_{\mu\neq 1} \left(\lambda_\mu^\alpha \hat{d}^\dagger_\mu-(\lambda^\alpha_\mu)^* \hat{d}_\mu\right)
\end{equation}
where $\dots$ contains fermionic bilinears that do not involve the zero energy mode or are exponentially suppressed interactions between Majorana fermions of the form $\hat{\gamma}_L^0\hat{\gamma}_R^0$. The coefficients $\lambda_\mu^\alpha$ are given by:
\begin{align}
    \lambda^\alpha_\mu= \frac{itg_B}{2}\sum_n \sigma e^{i\sigma \phi_{ext}/2}&( u^*_n \psi_\alpha^0(n+1)-v^*_{n+1,\mu} (\psi_\alpha^0(n))^*) + \sigma e^{-i\sigma \phi_{ext}/2}( v_{n,\mu}^* (\psi^0_\alpha(n+1))^* -u_{n+1,\mu}^*\psi_\alpha^0(n)
\end{align}
We emphasize that the operators $C_\alpha$ have a fermionic spatial support which is exponentially localized on each end of the ladder. Using the fact that $[\hat{\gamma}_\alpha^0,H_0]=0$, we can write the commutator in Eq.~\eqref{pertur-mf} as:
\begin{align}\label{commutator-pert}
    ig_B[\mathcal{D}_0,\hat{\gamma}_L^0]=\int_{-\infty}^{+\infty} dt \,F(\Delta_0 t) 2 C_L(t) (\hat a (t) + \hat a^\dagger (t)) 
\qquad \qquad 2 C_L(t) = e^{iH_0 t}[V_1,\hat{\gamma}^0_L] e^{-iH_0 t}    
\end{align}
with $\hat{a} (t) = \hat{a} e^{-i \omega_c t}$. The commutator $[V_1,\hat{\gamma}^0_L] = 2 C_L$ is exponentially localized on the left end of the ladder. $C_L(t)$ diffuses spatially as time evolves  but the time integral in Eq.~\eqref{commutator-pert} is cut off by $F(\Delta_0 t)$ and the result is again left-localized with exponential precision. 
Using the expression in Eq.~\eqref{eq:V1_expansion} for $C_\alpha$, we find:  
\begin{equation}
ig_B[\mathcal{D}_0,\hat{\gamma}_L^0]=-\frac{2}{\Delta_0}\sum_{\mu\neq 1} \lambda_{\mu}^L \hat{d}^\dagger_\mu\left[\hat{a}^\dagger\Tilde{F}\left(\frac{\epsilon_\mu +\omega_c}{\Delta_0}\right) +\hat{a} \Tilde{F}\left(\frac{\epsilon_\mu -\omega_c}{\Delta_0}\right)\right] -h.c.
\end{equation}
To make further progress, we need to determine the explicit form of $\tilde{F}(\omega)$, and there are two possibilities that can be distinguished:
\begin{enumerate}
    \item the single-photon energy $\hbar \omega_c$ does not cross the electronic spectrum. In that case, we introduce a minimal energy $\delta = {\rm min}_{\mu} |\omega_c - \varepsilon_\mu|$ and define $\tilde{F} (\omega)=-1/\omega$ down to $|\omega| \ge \delta/\Delta_0$. Since $\delta >0$, this choice is compatible with the desired exponential suppression of $F(\Delta_0 t)$ for large $t \simeq 1/\delta$. The smaller $\delta$, the larger the Majorana end states however expand spatially. The resulting Majorana polaritons are expressed as
    \begin{equation}\label{majorana-perturbative}
         \hat{\gamma}_\alpha \simeq\hat{\gamma}_\alpha^0 + 2 \sum_{\mu\neq 1}  \left( \frac{\lambda^\alpha_\mu \hat{d}^\dagger_\mu \hat{a}^\dagger +h.c.}{\epsilon_\mu + \omega_c} + \frac{\lambda^\alpha_\mu  \hat{d}_\mu^\dagger \hat{a} +h.c.}{\epsilon_\mu  - \omega_c} \right).
    \end{equation}    
    Reverting to the original fermionic operators, they adopt the form written in Eq.~\eqref{eq:maj_g} in the main text with the wavefunctions:
    \begin{align}
   & \Psi_L^{1+}(n) =\psi_L^{1+}(n) + \phi_L^{1+}(n)\; ,\qquad\Psi_L^{1-}(n) =\psi_L^{1-}(n) + \phi_L^{1-}(n)\; ,\\
       & \psi_L^{1+}(n)= 2\sum_{\mu\neq 1 }\frac{\lambda_{\mu}^L u_{n,\mu}}{\epsilon_{\mu}+\omega_c}, \qquad \qquad \psi_L^{1-} (n)= 2\sum_{\mu\neq 1} \frac{\lambda_\mu^L v_{n,\mu}}{\epsilon_{\mu}+\omega_c}, \\
       & \phi_L^{1+}(n)= 2\sum_{\mu\neq 1 }\frac{\lambda_{\mu}^L v_{n,\mu}^*}{\epsilon_{\mu}-\omega_c}, \qquad \qquad \phi_L^{1-} (n)= 2\sum_{\mu\neq 1} \frac{\lambda_\mu^L u_{n,\mu}^*}{\epsilon_{\mu}-\omega_c}. 
    \end{align}
    It can be checked numerically, as illustrated in Fig.~\ref{fig:maj_pol} of the main text, that these wavefunctions are exponentially localized on the left end of the ladder as expected from the quasi-adiabatic construction. A key feature of the hybrid Majorana operators in Eq.~\eqref{majorana-perturbative} is also that they commute with the full Hamiltonian,
    \begin{equation}
        [ \hat{H}, \hat{\gamma}_L] =  [ \hat{H}, \hat{\gamma}_R] = 0 
    \end{equation}
 up to exponentially small corrections. It implies a twofold degenerate spectrum where the even and odd parity sectors have the same energies. The Majorana operators are called {\bf strong edge modes}~\cite{Fendley_2012,jermyn2014,Alexandradinata_prb2016,Iemini_prl2017}.
    \item the single-photon energy crosses the energy of an electronic state, $\hbar \omega_c = \varepsilon_{\mu_0}$. In this case, the choice $\tilde{F} (\omega)=-1/\omega$ cannot be made for all $\omega$ without violating locality. We thus keep choosing $\tilde F(\omega) = -1/\omega$ except for energies $\epsilon_\mu$ close to $\epsilon_{\mu_0}$ where $\tilde F(\omega)$ has to be suppressed to ensure the exponential drop in real time. For simplicity, we use $\tilde F(\omega)=0$ only for $\omega=0$, and obtain
   \begin{equation}\label{majorana-perturbative2}
         \hat{\gamma}_r \simeq\hat{\gamma}_r^0 + 2 \sum_{\mu\neq 1,\mu_0}  \left( \frac{\lambda^r_\mu \hat{d}^\dagger_\mu \hat{a}^\dagger +h.c.}{\epsilon_\mu + \omega_c} + \frac{\lambda^r_\mu  \hat{d}_\mu  \hat{a}^\dagger +h.c.}{\epsilon_\mu  - \hbar \omega_c} \right).
    \end{equation}    
    where, in comparison with Eq.~\eqref{majorana-perturbative}, only the $\mu_0$ term has been excluded. In general, the commutation of the edge Majorana zero-modes with the Hamiltonian can be traced back to the choice $\tilde F(\omega) = -1/\omega$ made down to the smallest energies. Here, since there is one term missing, the Majorana polaritons no longer commute with the Hamiltonian which makes them {\bf weak edge modes}. The terms that are missing for commutation involve $\hat{d}_\mu a^\dagger$ and $\hat{d}_\mu^\dagger a$. Nevertheless, since these two operators have a vanishing expectation within the ground state manifold, it implies that $\hat{\gamma}_L$ and $\hat{\gamma}_R$ still commute with the Hamiltonian when restricted to the ground state manifold, as demonstrated in the main text in the broader non-perturbative context. Consequently, the spectrum remains twofold degenerate in the ground state but not for excited states.    
\end{enumerate}

Finally, it is important to note that the presence of weak edge modes directly arises from a resonance condition $\hbar \omega_c = \varepsilon_{\mu_0}$ between a single photon and an electronic excitation. For higher orders in $g$ in perturbation theory, we anticipate the possibility of multi-photon and multi-electron resonances. This will lift the degeneracy for most of the states in the many-body spectrum transforming the majorana modes into weak-edge modes. Indeed notice that in Fig. \ref{fig:stability} of the main text there is no resonance between the first excited state $\Delta E=\epsilon_2\simeq0.33$ and the cavity frequency $\omega_c=1$. 

\section{Perturbation theory}\label{sec:app_pert}
We apply now perturbation theory starting from the ground state with an applied magnetic field $\phi_0$ in the topological phase. In the following we will treat only the magnetic cavity case, calculations for the electric cavity can be done in the same way. The Hamiltonian can be expanded up to second order in $g_B$ as:
\begin{equation}
    H_B\simeq \omega_c \hat{a}^\dagger \hat{a} + H_0 + V_1 (\hat{a}+\hat{a}^\dagger) +(\hat{a}+\hat{a}^\dagger)^2V_2
\end{equation}
with 
\begin{align}
    &V_1= -t_0g_B\sum_{j,\sigma}i\sigma e^{i\sigma \phi_{ext}/2}\hat{c}^\dagger_{j,\sigma}\hat{c}_{j+1,\sigma} +h.c.\\
    &V_2= \frac{t_0 g_B^2}{2}\sum_{j,\sigma}e^{i\sigma\phi_{ext}/2}\hat{c}^\dagger_{j,\sigma}\hat{c}_{j,\sigma}
\end{align}
Both perturbations commute with the $Z_2$ symmetry which protects the ground state degeneracy, namely the number of fermions parity, and hence we can expect the degeneracy to be resilient to such perturbations. In particular we can treat the two sectors separetely. The ground states at $g_B=0$ are a product of cavity with no photons and the vacuum of bogoliubov quasiparticles, except for the non-local complex fermion $d_1$ which has $\epsilon_1\simeq 0$ and distinguishes between the two parity sectors. Note that the parity of the bogoliubov vacuum is not always the same and depends on Pfaffian of the Bogoliubov transformation \cite{majoranawf_Hegde_prb2016}. For simplicity let us assume that the even parity sector is the actual ground state, similar formulas can be derived in the other case. The ground states are then $\ket{0^{(0)}}=\ket{\emptyset}$ and $\ket{1^{(0)}}=d^\dagger_1\ket{0^{(0)}}$ where the superscript stands for zeroth order in $g$. The first order correction to the ground states is:
\begin{align}
    \ket{0^{(1)}} &= \ket{0^{(0)}}-\sum_{\mu\neq\nu\neq1}\frac{A^{\mu,\nu}}{\epsilon_\mu+\epsilon_\nu+\omega_c}\hat{d}^\dagger_\mu \hat{d}^\dagger_\nu \hat{a}^\dagger\ket{0^{(0)}}  - \sum_{\mu\neq1} \frac{A^{\mu,1}-A^{1,\mu}}{\epsilon_\mu +\omega_c}\hat{d}^\dagger_\mu \hat{a}^\dagger
    \ket{1^{(1)}}-\sum_\mu\frac{J_\mu}{\omega_c} \hat{a}^\dagger \ket{0^{(0)}}\\ \ket{1^{(1)}} &= \ket{1^{(0)}}-\sum_{\mu\neq\nu\neq 1}\frac{A^{\mu,\nu}}{\epsilon_\mu+\epsilon_\nu+\omega_c}\hat{d}^\dagger_\mu \hat{d}^\dagger_\nu \hat{a}^\dagger\ket{1^{(0)}} -\sum_{\mu\neq 1}\frac{B_\mu}{\epsilon_\mu+\omega_c}\hat{d}^\dagger_\mu \hat{a}^\dagger\ket{0^{(0)}} -\left(\frac{J'_1}{\omega_c} +\sum_{\mu\neq1} \frac{J_\mu}{\omega_c} \right)\hat{a}^\dagger \ket{1^{(0)}} 
\end{align}
where:
\begin{align}
    &A^{\mu,\nu}=\bra{0^{(0)}}\hat{d}_\nu \hat{d}_\mu \hat{a} \; V_1 (\hat{a}+\hat{a}^\dagger)\ket{0^{(0)}} =-ig_Bt_0\sum_{n}\sigma e^{i\sigma\phi_{ext}/2}u^*_{n,\mu}v^*_{n+1,\nu}-\sigma e^{-i\sigma\phi_0/2}u^*_{n+1,\mu}v^*_{n,\nu} \\
    &B_\mu=\bra{1^{(0)}}(\hat{d}^\dagger_1 \hat{d}_\mu+\hat{d}_\mu \hat{d}^\dagger_1 )\hat{a} \; V_1 (\hat{a}+\hat{a}^\dagger)\ket{1^{(0)}}=-ig_Bt_0\sum_n\sigma e^{i\sigma\phi_{ext}/2}(u^*_{n,\mu}u_{n+1,1}-v_{n,1}v^*_{n+1,\mu})+\\
    &\qquad\qquad\qquad\qquad\qquad\qquad\qquad\qquad\qquad-\sigma e^{-i\sigma\phi_0/2}(u^*_{n+1,\mu}u_{n,1}-v_{n+1,1}v^*_{n,\mu}) \\
    &J_\mu =\bra{0^{(0)}}\hat{d}_\mu \hat{d}^\dagger_\mu \hat{a} \; V_1 (\hat{a}+\hat{a}^\dagger)\ket{0^{(0)}}=-ig_Bt_0\sum_{n}\sigma e^{i\sigma\phi_{ext}/2}v_{n,\mu}v^*_{n+1,\mu}-h.c.\\
    &J'_1=\bra{1^{(0)}}\hat{d}^\dagger_1 \hat{d}_1 \hat{a} \; V_1 (\hat{a}+\hat{a}^\dagger)\ket{1^{(0)}}=-ig_Bt_0\sum_{n}\sigma e^{i\sigma\phi_{ext}/2}u^*_{n,1}u_{n+1,1}-h.c. \;\simeq J_1+O(e^{-L})\\
    &T_0= \bra{0^{(0)}} V_2 (\hat{a}+\hat{a}^\dagger)^2\ket{0^{(0)}}=\frac{g_B^2t_0}{2} \sum_ne^{i\sigma\phi_{ext}/2} \sum_\mu(v_{n,\mu}v^*_{n+1,\mu})+ h.c. \\
    &T_1=\bra{1^{(0)} }V_2 (\hat{a}+\hat{a}^\dagger)^2\ket{1^{(0)}}=\frac{g_B^2t_0}{2} \sum_ne^{i\sigma\phi_{ext}/2}\left(u_{n,1}^* u_{n+1,1}+ \sum_{\mu\neq 1}(v_{n,\mu}v^*_{n+1,\mu})\right)+ h.c.
\end{align}
With these it is also straightforward to calculate corrections to the ground state energies up to second order in $g$:
\begin{align}
    &E_0^{(2)}= E_0^{(0)} - \sum_{\mu>\nu \neq 1} \frac{|A^{\mu,\nu}-A^{\nu,\mu}|^2}{\epsilon_\mu+\epsilon_\nu+\omega_c} -\sum_{\mu\neq1}\frac{|A^{\mu,1}-A^{1,\mu}|^2}{\epsilon_\mu+\omega_c}-\frac{|\sum_{\mu} J_\mu|^2}{\omega_c} + T_0 \\
    &E_1^{(2)}= E_1^{(0)} - \sum_{\mu>\nu \neq 1} \frac{|A^{\mu,\nu}-A^{\nu,\mu}|^2}{\epsilon_\mu+\epsilon_\nu+\omega_c} -\sum_{\mu\neq1}\frac{|B_{\mu}|^2}{\epsilon_\mu+\omega_c}-\frac{|J_1' +\sum_{\mu\neq 1}J_\mu|^2}{\omega_c} +T_1
\end{align}
which give a renormalized gap:
\begin{equation}
    \Delta E^{(2)}_{gs}= \Delta E^{(0)}_{gs}-\frac{|J_1'|^2 - |J_1|^2}{\omega_c}-\sum_{\mu \neq 1} \frac{|B_{\mu}|^2 -|A^{\mu,1}-A^{1,\mu}|^2}{\epsilon_\mu +\omega_c} +T_1-T_0\;.
\end{equation}
Note that in principle this is not guaranteed to be exponentially small with system size, but can be checked to agree with the numerical results of Fig. \ref{fig:stability} in the main text.\\
Importantly we can connect perturbative expressions that we obtain by dressing the states to the polariton-Majorana operator. In particular one can check that:
\begin{equation}
\lambda^L_\mu=- \frac{A^{\mu,1} -A^{1,\mu}}{2}-\frac{B_\mu}{2}\qquad\lambda^R_{\mu} =i \,\frac{A^{\mu,1}+A^{1,\mu}}{2}  -i\frac{B_\mu}{2} 
\end{equation}
Hence the wavefunctions $\psi^{1+}_\alpha$ and $\psi_\alpha^{1-}$ can be extracted as connected matrix elements of polariton operators between the two ground state:
\begin{align}
    &\bra{0}\hat{c}_n \hat{a}\ket{1}_c=\bra{0}\hat{c}_n \hat{a}\ket{1}-\bra{0}\hat{c}_n\ket{1}\bra{0}\hat{a}\ket{0} \simeq-\sum_{\mu\neq 1} u_{n,\mu} \frac{B_\mu}{\epsilon_\mu +\omega_c} = \frac{1}{2}\left(\psi^{1+}_L -i\psi^{1+}_R\right)\\
    &\bra{0}\hat{c}_n^\dagger \hat{a}^\dagger \ket{1}_c=\bra{0}\hat{c}_n^\dagger \hat{a}^\dagger \ket{1}-\bra{0}\hat{c}_n^\dagger \ket{1}\bra{0}\hat{a}^\dagger \ket{0}\simeq -\sum_{\mu\neq 1} u^*_{n,\mu}  \frac{(A^{\mu,1})^*-(A^{1,\mu})^* }{\epsilon_\mu+\omega_c}=\frac{1}{2}\left((\psi^{1+}_L)^* -i(\psi^{1+}_R)^*\right)\\
    &\bra{0}\hat{c}_n^\dagger \hat{a}\ket{1}_c=\bra{0}\hat{c}_n^\dagger \hat{a}\ket{1}-\bra{0}\hat{c}_n^\dagger \ket{1}\bra{0} \hat{a}\ket{0}\simeq -\sum_{\mu\neq 1} v_{n,\mu} \frac{B_\mu}{\epsilon_\mu +\omega_c}= \frac{1}{2}\left(\psi^{1-}_L -i\psi^{1-}_R\right)\\
    &\bra{0}\hat{c}_n \hat{a}^\dagger\ket{1}_c=\bra{0}\hat{c}_n \hat{a}^\dagger\ket{1}-\bra{0}\hat{c}_n \ket{1}\bra{0}\hat{a}^\dagger\ket{0}\simeq -\sum_{\mu\neq 1} v^*_{n,\mu} \frac{(A^{\mu,1})^*-(A^{1,\mu})^*}{\epsilon_\mu +\omega_c}=\frac{1}{2}\left( (\psi^{1-}_L)^* -i(\psi^{1-}_R)^*\right)
\end{align}

\section{Details about the DMRG numerics}
\begin{figure}[!h]
    \centering
\includegraphics[width=0.8\textwidth]{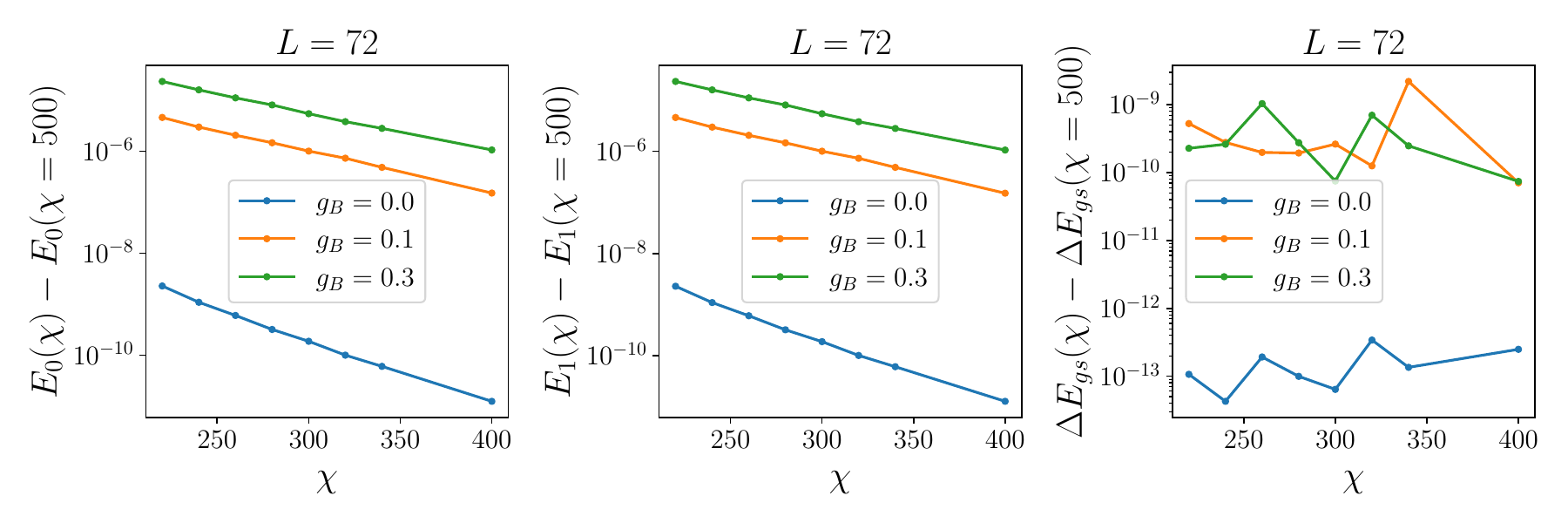}
    \caption{(a,b) Convergence of ground state energy for parity sector $p=0$ and $p=1$ for three values of the light-matter interaction strenght at $L=72$}
    \label{fig:dmrg}
\end{figure}

In order to perform our DMRG calculations for parity resolved ground and excited states we use the ITensor library \cite{itensor}. The MPS ansatz is mapped onto the physical degrees of freedom as in Ref \cite{bacciconi_arxiv2023} and we implement the fermionic parity quantum number. The photon Hilbert space truncated dimension is $N_{max}=64$, well above convergence. All of the results are obtained with a MPS bond dimension $\chi=400$, enough to obtain a discarded weight below $10^{-9}$ for the ground states and $10^{-7}$ for the excited states. The ground state degeneracy shown in the main text is also converged up to $10^{-8}$, even though the energy of each ground state itself do not reach this level of accuracy (Fig \ref{fig:dmrg}). This cancellation is rooted in the fact that the finite bond dimension affects in the same way both parity sector ground states.

\section{Mean-Field approximation}
In the following we briefly recap the mean-field treatment discussed in the text, which follows Ref. \cite{bacciconi_arxiv2023}. We will describe it for the electric field coupling but the same treatment can be done for the magnetic field. The starting point is to assume a separated light-matter ground state $\ket{\Psi}=\ket{\psi_{ph}}\ket{\psi_{m}}$. This leads to the decoupling of the hopping dressing as:
\begin{equation}
-t_\perp\sum_{j} e^{ig_E(\hat{a}+\hat{a}^\dagger)}  \hat{c}^\dagger_{+,j}\hat{c}_{-,j}\rightarrow  -t_\perp R\sum_j \left(e^{i\phi_E} \hat{c}^\dagger_{+,j}\hat{c}_{-,j}+h.c.\right) + T_\perp \cos( g_E(\hat{a}+\hat{a}^\dagger) ) + J_\perp \sin( g_E(\hat{a}+\hat{a}^\dagger) )
\end{equation}
where $Re^{i\phi_E}=\bra{\psi_{ph}}e^{ig_E(\hat{a}+\hat{a}^\dagger)} \ket{\psi_{ph}}$ and $T_\perp+iJ_\perp = \bra{\psi_{m}} \sum_j \hat{c}^\dagger_{j,+}\hat{c}_{j,-}\ket{\psi_{m}}$. Then in order to solve the decouple photon and matter problem we can do simple exact diagonalization on a truncated photon hilbert space and diagonalize the matter part by means of a Bogoliubov de-Gennes transformation. It is easy to check that the resulting fermionic spectrum is independent of $\phi_E$, signals that the coupling respect gauge invariance and photon condensation is forbidden. Solving matter and photon self-consistently up to convergence will give as a result a mean-field energy and a mean-field energy spectrum which can be constructed by just populating quasi-particles modes.

\end{document}